\documentclass[aps,showkeys,pra,showpacs,amssymb,amsmath]{revtex4}
\usepackage{amsmath}
\usepackage{graphicx}
\usepackage{dcolumn}
\usepackage{bm}
\usepackage[colorlinks=false,dvipdfm]{hyperref} 
\usepackage{setspace}

\begin{document}

\title{Maximal Quantum Violation of the CGLMP Inequality on Its Both Sides}

\author{Ming-Guang Hu}
\author{Dong-Ling Deng}
\author{Jing-Ling Chen}
\email{chenjl@nankai.edu.cn}
\address{Theoretical Physics Division, Chern Institute of Mathematics,
 Nankai University, Tianjin 300071, P. R. China\\PHONE:
 011+8622-2350-9287\\
FAX: 011+8622-2350-1532 }
\date{\today}

\begin{abstract}
\centerline{\textbf{Abstract}} We investigate the maximal violations
for both sides of the $d$-dimensional CGLMP inequality by using the
Bell operator method. It turns out that the maximal violations have
a decelerating increase as the dimension increases and tend to a
finite value at infinity.  The numerical values are given out up to
$d=10^6$ for positively maximal violations and $d=2\times 10^5$ for
negatively maximal violations. Counterintuitively, the negatively
maximal violations tend to be a little stronger than the positively
maximal violations. Further we show the states corresponding to
these maximal violations and compare them with the maximally
entangled states by utilizing entangled degree defined by von
Neumann entropy. It shows that their entangled degree tends to some
nonmaximal value as the dimension increases.
\end{abstract}

\maketitle

Bell inequality \cite{1964Bell}, arising from the
Einstein-Podolsky-Rosen (EPR) paradox \cite{1935Einstein}, has
developed into an effective criterion between the classical and
quantum theory. The famous Clauser-Horne-Shimony-Holt (CHSH)
inequality \cite{CHSH} for two entangled spin-1/2 particles has
always provided an excellent test-bed for experimental verification
of quantum mechanics against the predictions of local realism. In
2002, two research teams independently developed Bell inequalities
for high-dimensional systems: the first one is a Clauser-Horne type
(probability) inequality for two qutrits
\cite{2002Kaszlikowski,2002Chen}; the second one is a CHSH type
inequality for two qu$d$its ($d$ is an arbitrarily high dimension),
now known as the Collins-Gisin-Linden-Masser- Popoescu (CGLMP)
inequality \cite{2002Collins}. The tightness of the CGLMP inequality
was demonstrated in Ref. \cite{2002Lluis}. Such an inequality is
later modified into more compact forms \cite{2004Fu,2005Chen} and
its numerical violations for high-dimensional situations have also
been investigated in \cite{2006Chen,2008Gill}.

The CGLMP inequality for two $d$-dimensional systems is constrained
by two classical bounds at its two sides, one being positive and the
other negative \cite{2005Chen}. Both two classical bounds can be
violated by quantum behaviors. Such a property provides us the
chance to inspect the quantum nonlocality in the high-dimensional
cases. In principle, the maximally violating values of the
inequality can be achieved by searching all possible measurements on
all kinds of entangled states including mixed states. However, the
complexity of numerical simulation will increases in power as the
dimension increases, mainly because the number of variable
parameters during the searching process is proportional to $d^2+d$
(for pure states). In Ref. \cite{2006Chen}, Chen et al. have
investigated the maximal quantum violation of the CGLMP inequality
at its positive side by using the Bell operator method. Such a
method has a merit that it transforms the maximal violation problem
into a pure mathematical calculation of solving the maximal and
minimal eigenvalues of a matrix. By doing this, it needs a
prerequisite that a set of phase rules for fixing the measurement
projectors must be known so as to explicitly give out the matrix
elements. In general, the determination of these rules needs a large
quantity of numerical simulations. As it is, the rule for obtaining
the positively maximal violation of the CGLMP inequality has been
known before \cite{2006Chen,2002Collins}, but the one for the
negatively maximal violation is still unknown. Physically, it is
necessary for us to examine whether the negative violation is larger
than the positive side, since a stronger quantum nonlocality may be
indicated while the positive violation would be enough if that is
not the case.


In this paper, we first simplify the Bell operator method appeared
in \cite{2006Chen} and then give out, respectively, the phase rules
for obtaining the positively and negatively maximal violations of
the CGLMP inequality. After that, the matrix elements of Bell
operator are derived. Numerical results about maximal quantum
violations are achieved by solving the maximal and minimal
eigenvalues of the Bell operator. At last, the strength of quantum
nonlocality is compared between the positive and negative
violations, and their corresponding states are also discussed.
%

The CGLMP inequality as an extended version of the Bell-CHSH
inequality for $d$-dimensional Hilbert space has the form
\cite{2004Fu}
\begin{eqnarray}
I_d= Q_{11}+Q_{12}-Q_{21}+Q_{22}, \label{eq:Bell}
\end{eqnarray}
where $Q_{ij}$ are the Bell-type correlation functions defined by
probabilities in the way
\begin{eqnarray}
Q_{ij}= \frac{1}{S}\sum_{m=0}^{d-1}%
\sum_{n=0}^{d-1}f^{ij}(m,n)P(A_{i}=m,B_{j}=n),
 \label{Qij}
\end{eqnarray}
with the spin $S=(d-1)/2$ and eigenvalues of the correlation
function, $ f^{ij}(m,n)=S-M(\varepsilon (i-j)(m+n),d)$ [$\varepsilon
(x)=1$ for $x\geq0$ and $-1$ for $x<0$; $M(x,d)=x\;{\rm mod}\;d$.].
In classical theory, the Eq. (\ref{eq:Bell}) has two bounds
\cite{2005Chen} as
\begin{equation}\label{eq-bounds}
-2(\delta_{2d}+(1-\delta_{2d})\frac{d+1}{d-1})\leq I_d\leq 2,
\end{equation}
which can be violated in quantum theory. Following
\cite{2002Kaszlikowski,2002Collins,2002Acin,2001Durt,2000Kaszlikowski,2006Chen},
we use the unitary transformation setting of unbiased multiport beam
splitters (UMBS) \cite{1997Zukowski,2000Kaszlikowski} instead of a
complete $d$-dimensional unitary transformation to define projectors
[see Eq. (\ref{eq:prob})]. Unbiased $d$-port beam splitter (see Fig.
1) is a device with the following property: if a photon enters any
of the $d$ single input ports, its chances of exit are equally split
among the $d$ output ports. In fact one can always build the device
with the distinguishing trait that the elements of its unitary
transition matrix $T$ are solely powers of the root of unity $\gamma
=\exp(i2\pi/d)$, namely, $T_{kl}=\frac{1}{\sqrt{d}}\gamma^{kl}$. In
front of $i$th input port of the device a phase shifter is placed to
change the phase of the incoming photon by $\phi(i)$. These $d$
phase shifts, denoted for convenience as a ``vector" of phase shifts
$\hat{\phi}=(\phi(0),\phi(1),\cdots,\phi(d-1))$, are macroscopic
local parameters that can be changed by the observer. Therefore,
unbiased $d$-port beam splitter together with the $d$ phase shifters
perform the unitary transformation $U(\hat{\phi})$ with the entries
$U_{kl}=T_{kl}\exp[i \phi(l)]$. The diagram of such unitary
operation has been displayed in the Fig. 1. 

\begin{figure}
  \includegraphics[width=8.5cm]{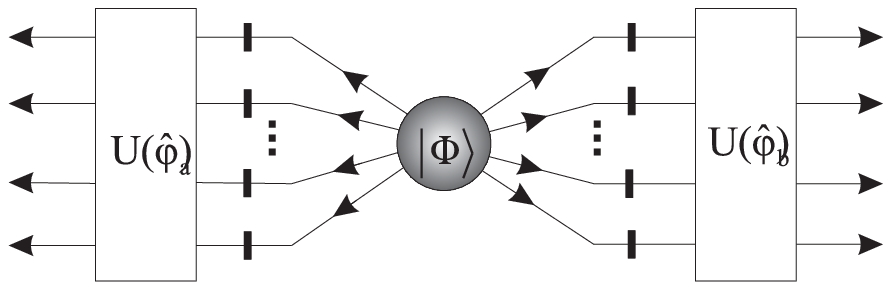}\\
  \caption{The diagram of unbiased multiport beam splitter. In front of the $i$th input port
of the device a phase shifter is placed to change the phase of the
incoming photon by $\phi(i)$. The whole action can be described by
unitary operator $U(\hat{\phi}_a)$ and
$U(\hat{\varphi}_b)$.}\label{fig1}
\end{figure}

An arbitrary entangled state of two qudits after Schmidt
decomposition reads
\begin{equation}\label{eq:state}
|\Phi\rangle=\sum_{i=0}^{d-1}\alpha_i|ii\rangle,
\end{equation}
with Schmidt coefficients $\{\alpha_i\}$ being real and satisfying
normalization condition $\sum_i|\alpha_i|^2=1$ \cite{2000Nielsen}.
One can see that if we take $\alpha_i=1/\sqrt{d}$ for all $i$ it is
just the maximally entangled state. The quantum prediction of the
joint probability $P(A_a=k,B_b=l)$ when $A_a$ and $B_b$ are measured
in the initial state $|\Phi\rangle$ is given by
\begin{eqnarray}
P(A_a=k,B_b=l)&=&\langle \Phi|\hat{P}(A_a=k)\otimes\hat{P}(B_b=l)|\Phi\rangle\nonumber\\
&=&{\rm Tr}\left[(U(\hat{\phi}_a)^{\dagger}\otimes
U(\hat{\varphi}_b)^{\dagger})\; \hat{\Pi}_k\otimes \hat{\Pi}_l\;
(U(\hat{\phi}_a)\otimes U(\hat{\varphi}_b))\;
|\Phi\rangle\langle\Phi|\right] \nonumber  \\
&=& \frac{1}{d^2}\sum_{j,m=0}^{d-1} \alpha_{j}\alpha_{m}
e^{i[\frac{2\pi}{d}(j-m)(k-l)+\phi_a(j)-\phi_a(m)+\varphi_b(j)-\varphi_b(m)]}\nonumber\\
&=& \frac{1}{d^2}\sum_{j,m=0}^{d-1} \alpha_{j}\alpha_{m}
\cos\left[\frac{2\pi}{d}(j-m)(k-l)+\phi_a(j)-\phi_a(m)+\varphi_b(j)-\varphi_b(m)\right],
\label{eq:prob}
\end{eqnarray}
where $\hat{\Pi}_k=|k\rangle\langle k|$,
$\hat{\Pi}_l=|l\rangle\langle l|$ are the projectors for observers
$A$ and $B$ not including UMBS transformation, respectively.
Substituting Eq. (\ref{eq:prob}) into the high-dimensional Bell
inequality (\ref{eq:Bell}), one gets the Bell expression for the
state $|\Phi\rangle$:
\begin{eqnarray}\label{eq-Id}
I_d(|\Phi\rangle)&=&\frac{1}{d}\sum_{j,m=0}^{d-1}\alpha_j\alpha_m\left[e^{i[\phi_1(j)-\phi_1(m)+\varphi_1(j)-\varphi_1(m)]}\sum_{k=0}^{d-1}\left(1-\frac{2k}{d-1}\right)e^{i\frac{2\pi}{d}k(j-m)}\right.\nonumber\\
&&+e^{i[\phi_1(j)-\phi_1(m)+\varphi_2(j)-\varphi_2(m)]}\sum_{k=0}^{d-1}\left(1-\frac{2k}{d-1}\right)e^{-i\frac{2\pi}{d}k(j-m)}\nonumber\\
&&-e^{i[\phi_2(j)-\phi_2(m)+\varphi_1(j)-\varphi_1(m)]}\sum_{k=0}^{d-1}\left(1-\frac{2k}{d-1}\right)e^{i\frac{2\pi}{d}k(j-m)}\nonumber\\
&&\left.+e^{i[\phi_2(j)-\phi_2(m)+\varphi_2(j)-\varphi_2(m)]}\sum_{k=0}^{d-1}\left(1-\frac{2k}{d-1}\right)e^{i\frac{2\pi}{d}k(j-m)}\right].
\end{eqnarray}
By using the summation formula
$$\sum_{k=0}^{d-1}\left(1-\frac{2k}{d-1}\right)e^{i\frac{2\pi}{d}k(j-m)}=\frac{2d}{(d-1)[1-e^{i\frac{2\pi}{d}(j-m)}]},\quad
(j\neq m)$$ we can reduce Eq. (\ref{eq-Id}) to a real-valued
expression, i.e.,
\begin{eqnarray}\label{eq-Id1}
I_d(|\Phi\rangle)&=&\frac{1}{d-1}\sum_{j\neq m}^{d-1}\alpha_j\alpha_m\frac{1}{\sin\left[\frac{\pi}{d}(j-m)\right]}\bigg\{-\sin\left[\phi_1(j)-\phi_1(m)+\varphi_1(j)-\varphi_1(m)-\frac{\pi}{d}(j-m)\right]\nonumber\\
&&+\sin\left[\phi_1(j)-\phi_1(m)+\varphi_2(j)-\varphi_2(m)+\frac{\pi}{d}(j-m)\right]\nonumber\\
&&+\sin\left[\phi_2(j)-\phi_2(m)+\varphi_1(j)-\varphi_1(m)-\frac{\pi}{d}(j-m)\right]\nonumber\\
&&-\sin\left[\phi_2(j)-\phi_2(m)+\varphi_2(j)-\varphi_2(m)-\frac{\pi}{d}(j-m)\right]\bigg\}.
\end{eqnarray}
From Eq. (\ref{eq-Id1}), we can see that $I_d$ depends on the state
coefficients $\{\alpha_i\}$ and phase shifts
$\{\phi_{1,2}(j),\varphi_{1,2}(m)\}$. For the positively maximal
values of $I_d$, there have been several investigations: when
$|\Phi\rangle$ is restricted in maximally entangled states, the
quantum violation was shown numerically in three dimension in
\cite{2002Kaszlikowski} and in an arbitrary dimension in
\cite{2002Collins}; when $|\Phi\rangle$ is an arbitrarily entangled
state, the maximal quantum violations were displayed in
\cite{2008Gill,2006Chen,2002Acin}. For both cases, the latter one
has a larger value of $I_d$ than the former and hence indicates a
stronger quantum nonlocality. Particularly, the approach developed
in Ref. \cite{2006Chen,2002Acin} is related to the Bell operator,
which can be defined by reparametrizing $I_d(|\Phi\rangle)$
according to
$$I_d(|\Phi\rangle)=\mathrm{Tr}(\hat{B}|\Phi\rangle\langle\Phi|)=\langle\Phi|\hat{B}|\Phi\rangle,$$
where $\hat{B}$ is the so-called Bell operator
\cite{1992Braunstein}. Under the bases $\{|jj\rangle\}$, the
non-vanishing elements of the Bell operator, $B_{jm}=\langle
jj|\hat{B}|mm\rangle$, from Eq. (\ref{eq-Id1}) are those
off-diagonal elements with
\begin{eqnarray}\label{eq-B}
B_{jm}|_{(j\neq m)}&=&\frac{1}{d-1}\frac{1}{\sin\left[\frac{\pi}{d}(j-m)\right]}\bigg\{-\sin\left[\phi_1(j)-\phi_1(m)+\varphi_1(j)-\varphi_1(m)-\frac{\pi}{d}(j-m)\right]\nonumber\\
&&+\sin\left[\phi_1(j)-\phi_1(m)+\varphi_2(j)-\varphi_2(m)+\frac{\pi}{d}(j-m)\right]\nonumber\\
&&+\sin\left[\phi_2(j)-\phi_2(m)+\varphi_1(j)-\varphi_1(m)-\frac{\pi}{d}(j-m)\right]\nonumber\\
&&-\sin\left[\phi_2(j)-\phi_2(m)+\varphi_2(j)-\varphi_2(m)-\frac{\pi}{d}(j-m)\right]\bigg\}.
\end{eqnarray}
If we know phases $\{\phi_{1,2}(j),\varphi_{1,2}(m)\}$, the whole
matrix $(B_{jm})$ then can be determined. Thus, the positively
maximal or negatively minimal quantum violations of Eq.
(\ref{eq-bounds}) would equal to the maximal or minimal eigenvalues
of the matrix $(B_{jm})$ and their corresponding eigenvectors are
the states $|\Phi\rangle$ that are capable to make the inequality
maximal or minimal. As a result, a physical problem of searching the
positively maximal or negatively minimal quantum violations of Eq.
(\ref{eq-bounds}) becomes a mathematical problem of solving the
maximal or minimal eigenvalues of the matrix $(B_{jm})$ for all
possible choices of phase shifts $\{\hat{\phi}_a,\hat{\varphi}_b\}$.

In order to determine the maximal violation at both hand sides of
the Bell inequalities (\ref{eq:Bell}), we must first choose such an
appropriate form of $\{\hat{\phi}_a,\hat{\varphi}_b\}$ as to $I_d$
can have maximal quantum violations. Just as pointed out in the Ref.
\cite{2002Acin,2001Durt,2006Chen}, the positively maximal quantum
values of $I_d$ can be achieved when one takes phase shifts by
\begin{equation}\label{eq:phi1}
\phi_1(j)=0,\quad \phi_2(j)=\frac{j\pi}{d},\quad
\varphi_1(j)=\frac{j\pi}{2d},\quad \varphi_2(j)=-\frac{j\pi}{2d}.
 \end{equation}
As to the negatively minimal values of $I_d$, we perform the
numerical search for the inequality and find that it can be achieved
by piecewise taking
\begin{subequations}
 \begin{align}\label{eq-phi2a}
\phi_1(j)=0,\quad\phi_2(j)=\frac{j\pi}{d},\quad\varphi_1(j)=\frac{j\pi}{2d},\quad\varphi_2(j)=-\frac{j\pi}{2d},
\end{align}
for $0\leq j\leq \left[\frac{d-2}{3}\right]$ (square brackets here
denote taking integer part value), and taking
\begin{align}\label{eq-phi2b}
\phi_1(j)=0,\quad \phi_2(j)=\frac{(d+j)\pi}{d},
\quad\varphi_1(j)=\frac{(d+j)\pi}{2d},\quad\varphi_2(j)=-\frac{(d+j)\pi}{2d},
\end{align}
for $\left[\frac{d-2}{3}\right]+1\leq j\leq
d-2-\left[\frac{d-2}{3}\right]$, and taking
\begin{align}\label{eq-phi2c}
\phi_1(j)=0,\quad\phi_2(j)=\frac{(2d+j)\pi}{d},\quad
\varphi_1(j)=\frac{(2d+j)\pi}{2d},\quad\varphi_2(j)=-\frac{(2d+j)\pi}{2d},
\end{align}
for $d-1-\left[\frac{d-2}{3}\right]\leq j\leq d-1$.
\end{subequations} From Eqn. (\ref{eq:phi1}), (\ref{eq-phi2a}), (\ref{eq-phi2b}) and
(\ref{eq-phi2c}), we can write the phase shifts into an uniform
format
\begin{eqnarray}\label{eq-phi}
\phi_1(j)=0,\phi_2(j)=\frac{n_j\pi}{d},\varphi_1(j)=\frac{n_j\pi}{2d},\varphi_2(j)=-\frac{n_j\pi}{2d},
 \end{eqnarray}
where $n_j$ take different values described above. The non-vanishing
matrix elements of $(B_{jm})$ under condition (\ref{eq-phi}) can be
expressed as
\begin{equation}\label{eq-B1}
B_{j,m}|_{(j\neq
m)}=\frac{4}{d-1}\frac{\sin[(2(j-m)-(n_j-n_m))\pi/2d]}{\sin[(j-m)\pi/d]}.
\end{equation}
When $n_j=j$, the Eq. (\ref{eq-B1}) returns to
$B_{j,m}=2(1-\delta_{jm})/(d-1)\cos[(j-m)\pi/2d]$ appeared in
\cite{2006Chen}. Now that matrix elements of $(B_{jm})$ are all
known, the remainder is to get the extreme eigenvalues. As follows,
we apply the so-called power method in the numerical computing field
to achieve such task. At the same time the normalized eigenvectors
corresponding to these extremum are just Schmidt coefficients
$\{\alpha_i\}$ of Eq. (\ref{eq:state}).

It is natural to ask about the optimality of the chosen set of
measurements.  Certainly, the maximal and minimal values obtained by
using these phase settings are only when the unitary transformation
is restricted in the experimental settings of UMBS. For a general
case, we have also performed a numerical search for the inequality,
by using universal $SU(d)$ transformation instead of the above UMBS
settings up to $d=9$. The maximal and minimal values of $I_d$ are in
accordance with the results obtained by the above configuration and
the numerical values have been shown in Fig. 2.

In order to get a more absolute measure of nonlocality for
high-dimensional Bell inequalities, let us consider the initial
entangled state (\ref{eq:state}) mixed with some amount of noise
\cite{2000Kaszlikowski}
\begin{equation}\label{eq:mixsta}
\rho(F)=F\rho_{\text{noise}}+(1-F)|\Phi\rangle\langle\Phi|,
\end{equation}
where the positive parameter $F\leq 1$ determines the ``noise
fraction" within the full state and
$\rho_{\text{noise}}=\mathbf{1}/d^2$ (the bold $\mathbf{1}$ is a
$d\times d$ unit matrix). The threshold minimal $F_{\min}$, for
which the state $\rho(F)$ still allows a local realistic model, will
be our numerical value of the strength of violation of local realism
by the quantum state $|\Phi\rangle$. It has an indication that the
higher $F_{\min}$ is, the more noise admixture will be required to
hide the nonclassicality of the quantum prediction, and so the
stronger the quantum nonlocality is. For both the right-hand side
and the left-hand side of Eq. (\ref{eq-bounds}), we have the
relations between $F_{\min}$ and the maximal violations
$(I_d^{\max})_\Phi$ for pure states
\begin{eqnarray}
F_{\min}^{\text{right}}&=&1-2/I_d^{\max}(|\Phi\rangle),\\
F_{\min}^{\text{left}}&=&1+2\left[\delta_{2d}+(1-\delta_{2d})\frac{d+1}{d-1}\right]/I_d^{\min}(|\Phi\rangle).\nonumber
\end{eqnarray}

As for high-dimensional Bell inequalities, the states $|\Phi\rangle$
corresponding to maximal violations are generally not the maximal
entangled states any more. In order to measure the deviation of
$|\Phi\rangle$ from maximal entangled states by their entangled
degree, we use the von Neumann entropy of the reduced pure states
(party $A$ for example)
$\rho_A=\mathrm{Tr}_B(|\Phi\rangle\langle\Phi|)$ \cite{2000Nielsen}
which quantifies the bi-particle entanglement and is defined by
\begin{equation}\label{eq:vonN}
S(\rho_A)=-\mathrm{Tr}(\rho_A\log\rho_A)=-\sum_{i=0}^{d-1}\alpha_i^2\log\alpha_i^2,
\end{equation}
where the logarithms are taken to base two. For a pure state
$\rho_{AB}$ of two subsystems $A$ and $B$, the von Neumann entropy
of the reduced density operators Eq. (\ref{eq:vonN}) can effectively
distinguish classical and quantum mechanical correlations, i.e.,
quantify entanglement \cite{1997Vedral}. 
For the maximal entangled pure state (or GHZ state)
$|\psi_G\rangle=\frac{1}{\sqrt{d}}\sum_{j=0}^{d-1}|jj\rangle$, the
Eq. (\ref{eq:vonN}) for such state reads $S_{\max}=
S(\rho_A)=-\sum_{j=0}^{d-1}(1/d)\log(1/d)=\log d$ that corresponds
to the maximal entropy $\log d$ in a $d$-dimensional Hilbert space.
So we can use the rate $S(\rho_A)/S_{\max}$ as a useful criterion to
measure the deviation; the smaller the rate is, the bigger the
deviation is.

\begin{figure}
  \includegraphics[width=0.6\columnwidth]{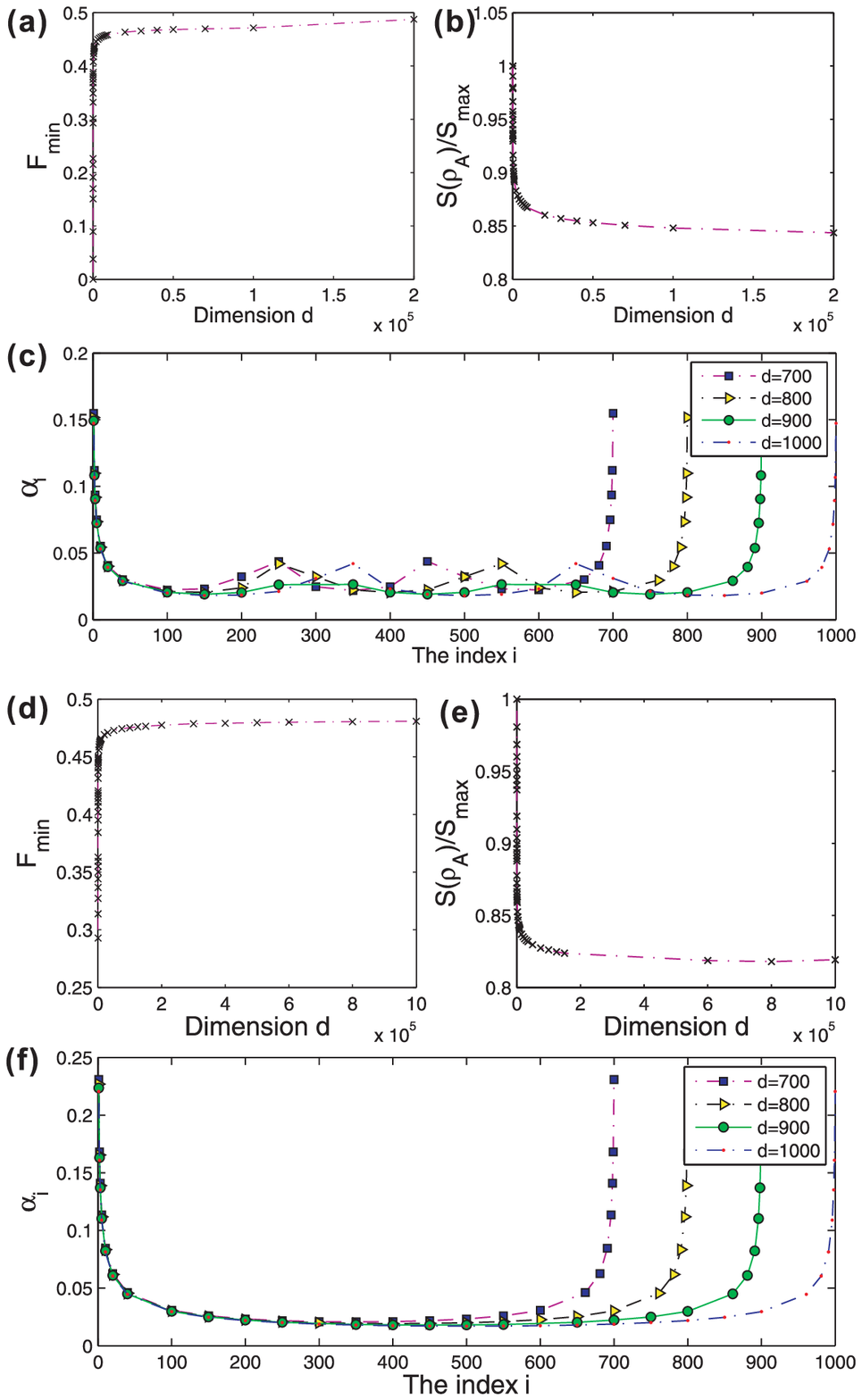}\\
  \caption{(Color online)The violation illustration of the CGLMP inequality in the
  $d$-dimensional Hilbert space. (a), (b), (c) describe the negative maximal violations,
  while (d), (e), (f) describe the positive
  maximal violations.}\label{fig3}
\end{figure}

In Fig. 2 we show numerically the maximal violations up to $d=10^6$
for positively maximal violations and $d=2\times 10^5$ for
negatively maximal violations. It is seen from Fig. 2(a)(d) that the
strength of quantum nonlocality denoted by $F_{\min}$ have a
decelerating increase with the growing system dimension $d$ and it
approaches a finite maximal value when $d$ goes to infinity. The
increasing trend also shows a stronger quantum nonlocality in a
higher dimensional system. An interesting phenomenon is that
although the quantum nonlocality becomes stronger in a higher
dimensional system, the corresponding state, from Fig 2.(b)(e), has
a decreasing entangled degree and hence deviates farther from the
most entangled state. From Fig. 2(c)(f), these Schmidt coefficients
of states corresponding those maximal quantum violations have a
symmetric distribution and they obey the same pattern no matter how
big the dimension is. By comparing Fig. 2(a-c) with Fig. 2(d-f),
there are several differences between the positively and negatively
maximal violations: (i) at the same dimension, the negatively
maximal violation of $I_d$ has a little larger $F_{\min}$ than the
positively maximal violation (e.g. when $d=2\times 10^5$, the
negatively maximal violation has $F_{\min}=0.487$ while the
positively maximal violation has $F_{\min}=0.477$); (ii) compared
with the positively maximal side, there is a smaller deviation for
the states corresponding to negatively maximal side (e.g. when
$d=1\times 10^5$, the negatively maximal violation has
$S(\rho_A)/S_{\max}=0.848$ while the positively maximal violation
has $S(\rho_A)/S_{\max}=0.826$). Note that for $d=3$, the negative
side of $I_d$ has no violation with $F_{\min}=0$ and so at this
point it does not satisfy the above general rules.

In summary, continued with the previous paper \cite{2006Chen}, we
have investigated the maximal violations of the CGLMP inequality for
two $d$-dimensional systems further. The maximal violation occurs at
the nonmaximally entangled state, which is the eigenvector of Bell
operator with the maximal eigenvalue. Numerically, we display the
maximal violating values up to $d=10^6$ for the positively maximal
violation and $d=2\times 10^5$ for the negatively maximal violation.
It gives us a straightforward picture on the trend of quantum
nonlocality with the dimension of systems increasing. That is, the
maximal violations increase as the dimension $d$ grows and
approaches a finite value when $d$ becomes infinity. In addition,
deviations of the entangled pure states that are in accordance with
maximal violations from maximally entangled states have also been
discussed. It is shown that as the dimension $d$ increases, the
deviation also approaches some constant limit.

Particulary, by comparing the violations at both sides of $I_d$, we
show that the negatively maximal violation has a little stronger
quantum nonlocality than the positively maximal violation has. This
may be counterintuitive to the previous notion that the positively
maximal violation of the CGLMP inequality is enough. Also, by giving
out the phase rules and the elements of the Bell operator matrix, we
have in fact given a convenient way to calculate the maximal
violation of the CGLMP inequality on its both sides. It can serve as
a useful reference for experimental setting when testing the CGLMP
inequality. The experimental test has been performed to verify the
CGLMP inequality for the first few high-dimensional systems
\cite{2002Vaziri}, and it shows that indeed there exist nonmaximally
entangled states violating the inequality more strongly than the
maximal entangled ones . Thus, the nonmaximally entangled states in
high-dimensional quantum systems as discussed in this paper may turn
to be comparably useful when applied to quantum cryptography and
quantum communication complexity
\cite{1991Ekert,2003Kasz,2004Durt,2002Brukner,2004Brukner}, in which
previous researches are mostly based on maximally entangled states.

\begin{acknowledgements}
This work is supported in part by NSF of China (Grants No. 10575053
and No. 10605013), Program for New Century Excellent Talents in
University, and the Project-sponsored by SRF for ROCS, SEM.
\end{acknowledgements}

\newpage
\section*{Figure legends}
\textbf{Figure 1}: The diagram of unbiased multiport beam splitter.
In front of the $i$th input port of the device a phase shifter is
placed to change the phase of the incoming photon by $\phi(i)$. The
whole action can be described by unitary operator $U(\hat{\phi}_a)$
and $U(\hat{\varphi}_b)$.

\textbf{Figure 2}: (Color online)The violation illustration of the
CGLMP inequality in the $d$-dimensional Hilbert space. (a), (b), (c)
describe the negative maximal violations, while (d), (e), (f)
describe the positive maximal violations.

\newpage
\section*{Figures}
\begin{figure}[tbph]
  \centering  \label{fig1}
  \includegraphics[width=0.8\columnwidth]{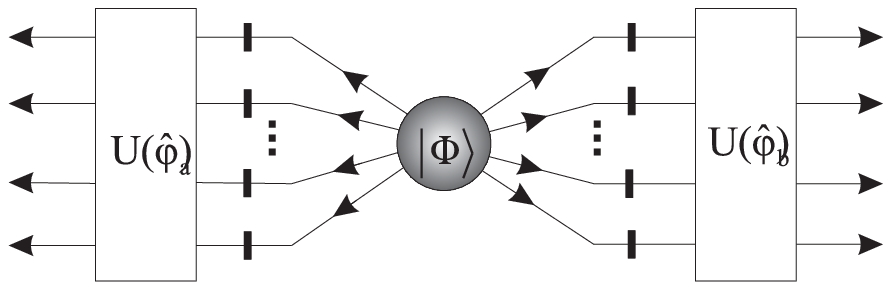}
  \\Figure 1\\
\end{figure}

\begin{figure}[tbph]
  \centering  \label{fig2}
  \includegraphics[width=0.7\columnwidth]{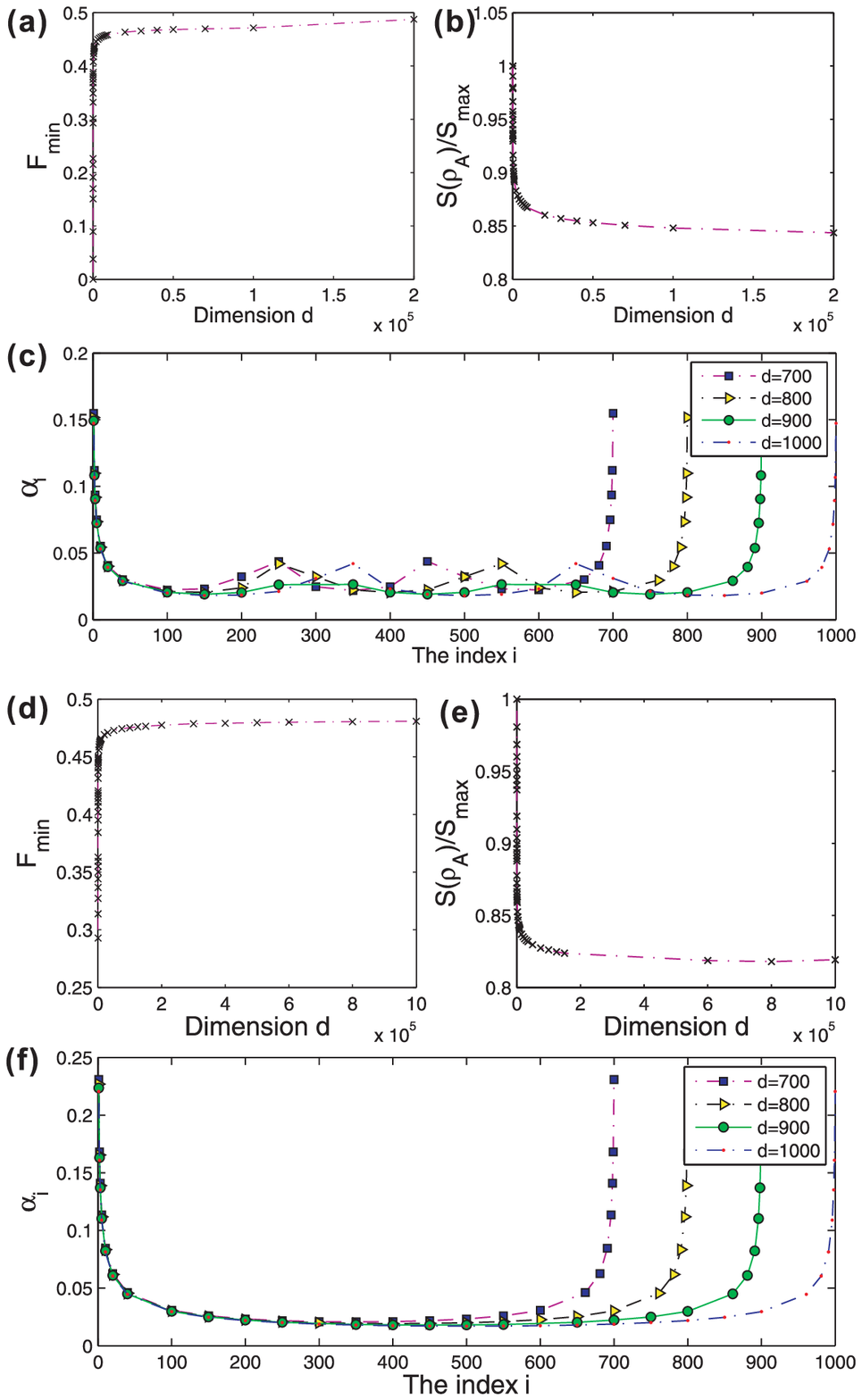}
  \\Figure 2\\
\end{figure}


%

\begin{thebibliography}{30}
\bibitem{1964Bell}
J. S. Bell, Physics (Long Island City, N.Y.) 1 (1964) 195.
\bibitem{1935Einstein}
A. Einstein, B. Podolsky, N. Rosen,  Phys. Rev. 47 (1935) 777.
\bibitem{CHSH}
J. F. Clause, M. A. Horne, A. Shimony, R. A. Holt, Phys. Rev. Lett.
23 (1969) 880.
\bibitem{2002Kaszlikowski}
D. Kaszlikowski, L. C. Kwek, J. L. Chen, M. \.{Z}ukowski, C. H. Oh,
Phys. Rev. A  65 (2002) 032118.
\bibitem{2002Chen}
J. L. Chen, D. Kaszlikowski, L. C. Kwek, C. H. Oh, Mod. Phys. Lett.
A 17 (2002) 2231.
\bibitem{2002Collins}
D. Collins, N. Gisin, N. Linden, S. Massar, S. Popescu, Phys. Rev.
Lett. 88 (2002) 040404.
\bibitem{2002Lluis}
Lluis Masanes, Quantum Inf. Comput. 3 (2002) 345.
\bibitem{2004Fu}
L. B. Fu, Phys. Rev. Lett. 92 (2004) 130404.
\bibitem{2005Chen}
J. L. Chen, C. Wu, L. C. Kwek, D. Kaszlikowski, M. \.{Z}ukowski, and
C. H. Oh, Phys. Rev. A 71 (2005) 032107.
\bibitem{2006Chen}
J. L. Chen, C. Wu, L. C. Kwek, C. H. Oh, M. L. Ge, Phys. Rev. A 74
(2006) 032106.
\bibitem{2008Gill}
 S. Zohren and R. D. Gill, Phys. Rev. Lett. 100 (2008) 120406.
\bibitem{1997Zukowski}
M. \.{Z}ukowski, A. Zeilinger, M. A. Horne, Phys. Rev. A 55 (1997)
2564.
\bibitem{2000Kaszlikowski}
D. Kaszlikowski, P. Gnaci\'{n}ski, M. \.{Z}ukowski, W. Miklaszewski,
A. Zeilinger, Phys. Rev. Lett. 85 (2000) 4418.
\bibitem{2002Acin}
A. Ac\'{i}n, T. Durt, N. Gisin, J. I. Latorre, Phys. Rev. A 65
(2002) 052325.
\bibitem{1992Braunstein}
S. L. Braunstein, A. Mann, M. Revzen, Phys. Rev. Lett. \textbf{68}
(1992) 3259.
\bibitem{2001Durt}
T. Durt, D. Kaszlikowski, M. \.{Z}ukowski, Phys. Rev. A 64 (2001)
024101.
\bibitem{2000Nielsen}
M. A. Nielsen, I. L. Chuang, \emph{Quantum Computation and Quantum
Information} (Cambridge University Press, Cambridge, 2000).
\bibitem{1997Vedral}
V. Vedral, M. B. Plenio, M. A. Rippin, P. L. Knight, Phys. Rev.
Lett. 78 (1997) 2275.
\bibitem{2002Vaziri}
A. Vaziri, G. Weihs, A. Zeilinger, Phys. Rev. Lett. 89 (2002)
240401.
\bibitem{1991Ekert}
A. K. Ekert, Phys. Rev. Lett. 67 (1991) 661.
\bibitem{2003Kasz}
D. Kaszlikowski, D. K. L. Oi, M. Christandl, K. Chang, A. Ekert, L.
C. Kwek, C. H. Oh, Phys. Rev. A 67 (2003) 012310.
\bibitem{2004Durt}
T. Durt, D. Kaszlikowski, J. L. Chen, L. C. Kwek, Phys. Rev. A 69
(2004) 032313.
\bibitem{2002Brukner}
C. Brukner, M. \.{Z}ukowski, A. Zeilinger, Phys. Rev. Lett. 89
(2002) 197901.
\bibitem{2004Brukner}
C. Brukner, M. \.{Z}ukowski, J.-W. Pan, A. Zeilinger, Phys. Rev.
Lett. 92 (2004) 127901.

%
%
%

\end{thebibliography}
\end{document}